\documentclass[twocolumn,showpacs,preprintnumbers,amsmath,amssymb]{revtex4}

\usepackage{graphicx}
\usepackage{dcolumn}
\usepackage{bm}

\usepackage{amsthm}

\theoremstyle{plain}
 
\newtheorem*{lemma}{Lemma}

\newtheorem*{main}{Theorem}

\font\tenscr=rsfs10 scaled1100
\font\sevenscr=rsfs7 
\font\fivescr=rsfs5 
\skewchar\tenscr='177 
\skewchar\sevenscr='177
\skewchar\fivescr='177 
\newfam\scrfam 
\textfont\scrfam=\tenscr
\scriptfont\scrfam=\sevenscr 
\scriptscriptfont\scrfam=\fivescr

\def\scri{{\fam\scrfam I}} 

\def\O{\mathcal{O}}

\begin{document}



\title{On the nonexistence of conformally flat slices in the Kerr and other stationary spacetimes}

\author{Juan Antonio Valiente Kroon}
 \email{jav@ap.univie.ac.at}
\affiliation{
Institut f\"ur Theoretische Physik der Universit\"at Wien,
Boltzmanngasse 5,
1090 Wien,
Austria. }

\date{\today}

\begin{abstract}
It is proved that a stationary solutions to the vacuum Einstein field
equations with non-vanishing angular momentum have no Cauchy slice
that is maximal, conformally flat, and non-boosted. The proof is based
on results coming from a certain type of asymptotic expansions near
null and spatial infinity ---which also show that the developments of
Bowen-York type of data cannot have a development admitting a smooth
null infinity---, and from the fact that stationary solutions do admit a
smooth null infinity.
\end{abstract}

\maketitle

\emph{Introduction.-} One of the basic tenets of what R. Penrose has
called the establishment's view concerning the Physics of black holes, is
that an asymptotically flat spacetime containing black holes will
eventually settle down to a Kerr black hole. This point of view has
been supported by the \emph{no hair} theorems of Israel, Carter and
Robinson \cite{Isr67,Car71,Rob75b}, and confirmed by numerical
simulations and perturbative analysis of the evolution of black hole
initial data with non-vanishing angular momentum ---see
e.g. \cite{BraSei95,BakBruCamLouTak01,DaiLouTak02,GleNicPriPul98}. A
great majority of these numerical simulations have made use of initial
data sets which are conformally flat, and have a second fundamental
form given by the so-called Bowen-York Ansatz \cite{BowYor80}. As
noted by several authors, the Bowen-York data do not reduce to 
Kerr data under any choice of the parameters. Now, there are several
reasons why it would be of interest to have slices of the Kerr
spacetime that are conformally flat. For example, one would like to
analyse the evolution of a pair of initially very close spinning
black holes as a perturbation of the Kerr solution, and not as a
perturbation of Schwarzschild as it has been done so far ---see
e.g. \cite{GleNicPriPul98}.

Garat \& Price \cite{GarPri00} have shown that the Kerr spacetime has
no axially symmetric, conformally flat slices that smoothly approach to the
standard Schwarzschildean slices of constant time as the angular
momentum goes to zero. Their argument was based on a perturbative
analysis of the Cotton-Bach tensor ---which locally characterises
conformally flat 3-dimensional hypersurfaces--- of the prospective
slices up to second order in the angular momentum. Nevertheless, as
noted by the authors themselves, there is, in principle, room for the
existence more exotic slices than the ones covered by their Ansatz.

This letter presents a theorem that precludes the existence of
maximal, non-boosted, conformally flat slices not only in the Kerr
spacetime, but also in any stationary spacetime with non-vanishing
angular momentum. The proof is based in general conformal properties
of the stationary solutions to the Einstein field equations, and makes
use not only of the information contained in the slices, but also of
features concerning their developments.

Let $\widetilde{S}$ be a Cauchy hypersurface, and let
$\widetilde{h}_{\alpha\beta}$, $\widetilde{\chi}_{\alpha\beta}$ be,
respectively, its first and second fundamental forms. We shall write
$\widetilde{\chi}=\widetilde{\chi}_{\gamma}^{\phantom{\gamma}\gamma}$. The
hypersurface $\widetilde{S}$ could have several asymptotically flat
regions. We concentrate our attention on one of them. Let
$\{y^\alpha\}$ be coordinates in this asymptotic region such that,
\[
\widetilde{h}_{\alpha\beta}=\left(1+\frac{2m}{|y|}\right)\delta_{\alpha\beta}+\O\left(\frac{1}{|y|^2}\right), \qquad \widetilde{\chi}_{\alpha\beta}=\O\left(\frac{1}{|y|^2}\right),
\]
as $|y|\rightarrow\infty$, where $m$ is the ADM mass associated with
this end. The (ADM) linear and angular momentum of the asymptotically
flat region are defined by,
\begin{eqnarray*}
&& P_\alpha=\frac{1}{8\pi}\lim_{r\rightarrow\infty}\int_{S_r}(\widetilde{\chi}_{\alpha\beta}-
\widetilde{\chi}\widetilde{h}_{\alpha\beta})n^\beta dS_r, \\
&& J_\alpha=\frac{1}{8\pi}\lim_{r\rightarrow\infty}\int_{S_r}\epsilon_{\alpha\beta\gamma}y^{\beta}(\widetilde{\chi}^{\gamma\rho}-\widetilde{\chi}\widetilde{h}^{\gamma\rho})n_\rho dS_r,
\end{eqnarray*}
respectively, where $S_r$ denotes the 2-sphere $\{|y|=r\}$, and
$n^\alpha$ is the outward unit normal.

Our main theorem is the following:

\begin{main}
Let
$(\widetilde{S},\widetilde{h}_{\alpha\beta},\widetilde{\chi}_{\alpha\beta})$
be a maximal, asymptotically Euclidean,
conformally flat initial data set for the Einstein vacuum field
equations. Assume, also, that the initial data set contains no linear
momentum, and that the conformally rescaled second fundamental form
admits an expansion of the form given by equation (\ref{conformal_2nd}). 
If, furthermore, the spacetime is stationary, then the angular momentum
of the data must vanish.
\end{main}

Thus, stationary spacetimes ---whence the Kerr solution--- admit no
maximal, conformally flat, non-boosted slices with a second
fundamental form of the sort prescribed by equation
(\ref{conformal_2nd}) if their ADM angular momentum is non-zero. If
the angular momentum were non-zero, then the Schwarzschild solution
would render a counterexample to our claim.

Some remarks concerning the hypothesis of the theorem come into
place. Firstly, the assumption of maximal slices ---i.e. such that
$\widetilde{\chi}$--- is natural, for very little is know about how to
solve the Einstein constraint equations for non-maximal
data. Secondly, the slices have been assumed to be non-boosted, for
the presence of linear momentum leads to non-smooth solutions of the
Hamiltonian constraint ---see \cite{DaiFri01}. Existence of solutions
to the Hamiltonian constraint for a conformal second fundamental form
of the sort given by equation (\ref{conformal_2nd}) ---under some
extra conditions--- has been discussed in the aforementioned
reference.

\emph{Sketch of the proof:} the proof of the theorem relies heavily on
some conformal properties of stationary spacetimes and, in particular,
on features of a certain kind of asymptotic expansions near spatial
infinity that can be obtained from a representation of spatial
infinity introduced by Friedrich \cite{Fri98a}. For asymptotically
Euclidean, conformally flat data with a second fundamental form like the
one given in equation (\ref{conformal_2nd}) ---no stationarity being
assumed here--- , the asymptotic expansions contain  certain
logarithmic terms which preclude the existence of a smooth null
infinity. These logarithmic terms vanish, if and only if, the angular
momentum of the data vanishes. It can be seen that this implies that
the development of such data admits no smooth null infinity. On the
other hand, it is known that the developments of stationary data do
admit a smooth null infinity. Consequently, if the initial data is to
be stationary, its angular momentum must vanish. \hfill $\Box$

\medskip
There are, nevertheless, a number of technical issues that one has to
analyse. These will be discussed in the sequel. In particular, there
is the fact that the gauge in which the expansions near spatial and
null infinity are obtained is different to that in which the existence
of a smooth null infinity for stationary spacetimes has been
proved. It could well be the case that the non-smoothness in the
expansions is a product of a bad choice of gauge. As it turns out,
this is not the case, and the way one can do it is by comparing them
in a gauge well adapted to null infinity ---the NP gauge.

\bigskip
\emph{Expansions near spatial and null infinity.-} Penrose
\cite{Pen63} has introduced the seminal idea of studying the
spacetimes describing isolated systems by means of a manifold with
boundary --\emph{the unphysical spacetime}--- $(M,g_{\mu\nu})$. This
unphysical spacetime is obtained from the original one,
$(\widetilde{M},\widetilde{g}_{\mu\nu})$ by means of a conformal
rescaling $g_{\mu\nu}=\Omega^2\widetilde{g}_{\mu\nu}$. The boundary of
the unphysical spacetime consists, at least, of the so-called null
infinity $\scri=\scri^+\cup\scri^-$, and of a point, $i^0$, called
spatial infinity. The spacetime
$(\widetilde{M},\widetilde{g}_{\mu\nu})$ will be said to admit a
smooth null infinity, if a conformal compactification can be
introduced in such a way that the resulting null infinity is a smooth
submanifold.

Friedrich has introduced a certain representation of spatial infinity
as a cylinder ---the cylinder at spatial infinity, $I$
\cite{Fri98a}. A major feature of this representation of null
infinity, is that the location and structure of the conformal boundary
of the spacetime is known a priori, if the Einstein constraint
equations have been solved. We shall refer to the gauge leading to
this representation of spatial infinity as to the F-gauge. The
equations governing the unphysical spacetime ---the Conformal Einstein
field equations--- reduce, upon evaluation on $I$, to an interior
system of transport equations.  This system of transport equations
enables us to obtain a certain type of asymptotic expansions, which in
turn allow to connect properties of the spacetime at null
infinity, with properties of the initial data. In particular, the
components $\phi_j$, $j=0,\ldots,4$ of the \emph{rescaled Weyl spinor}
$\phi_{abcd}$ can be expanded as,
\begin{equation}
\phi_{j}\sim \sum_{p\geq 0}\frac{1}{p!}\phi_j^{(p)}(\tau,\theta,\varphi)\rho^p, \label{expansion}
\end{equation}
$j=0,\ldots,4$ where $\tau$ is a coordinate such that the locus
of null infinity is given by $\tau=\pm 1$, and $(\theta,\varphi)$ are
the standard polar coordinates. The cylinder at spatial infinity
corresponds to the points for which $\rho=0$ and $-1\leq \tau \leq 1$.
The coefficients $\phi_j^{(p)}(\tau,\theta,\varphi)$ are determined by
the interior system at $I$.

\begin{figure}[t]
\centering
\includegraphics[width=.4\textwidth]{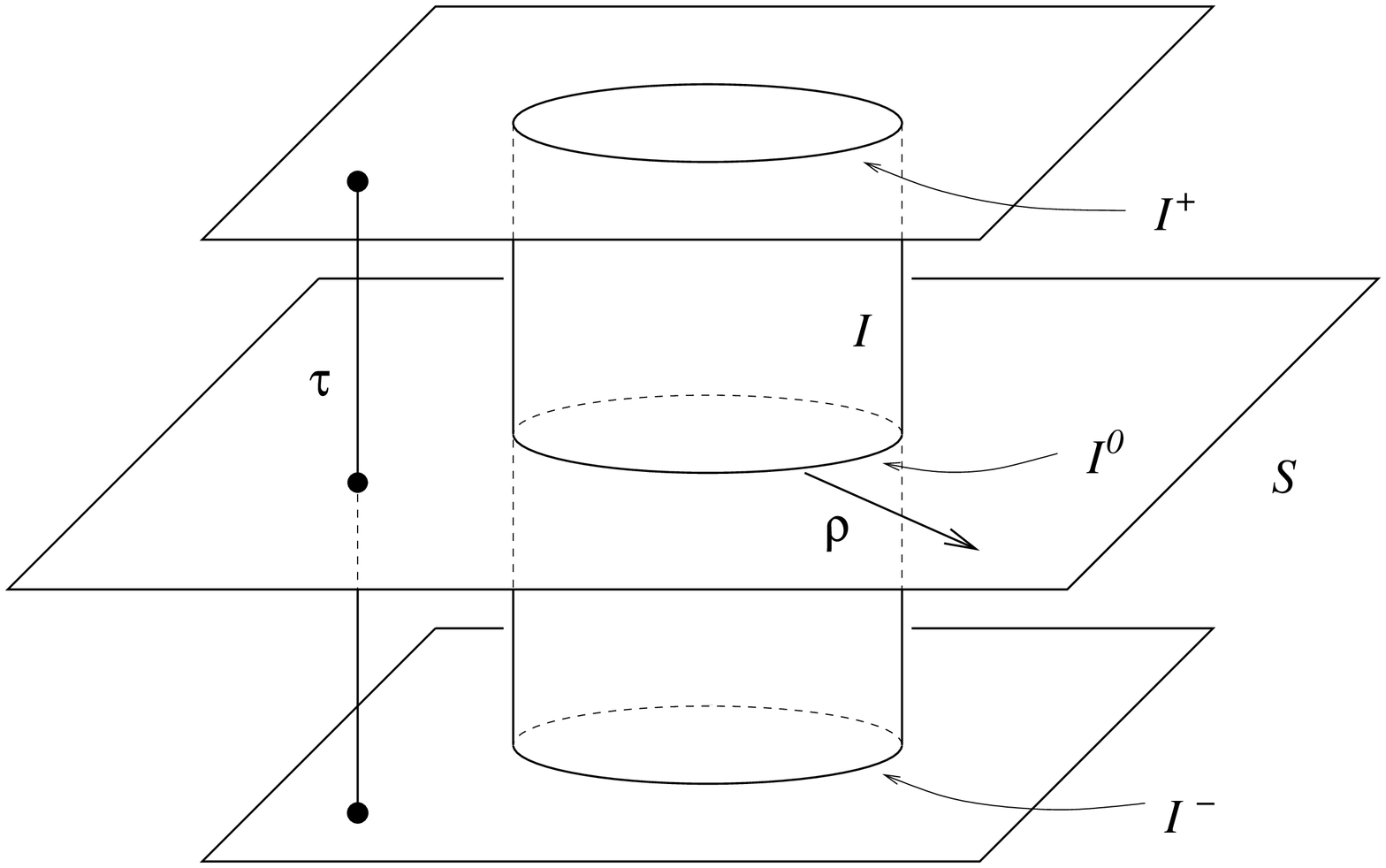}
\put(-25,120){$\scri^+$}
\put(-25,25){$\scri^-$}
\caption[]{}
\label{new:fig}
\caption{The region of the ---unphysical spacetime--- near spatial and null infinities. The conformal factor leading to this representation is given by $\Theta=\omega\vartheta^{-2}(1-\tau^2)$ with $\omega=\vartheta/|d\vartheta|$.}
\end{figure}

A set of scripts in the computer
algebra system {\tt Maple V} have been written in order to calculate the
aforementioned expansions. They have been used to analyse the
behaviour near spatial infinity of time symmetric solutions
\cite{Val03b,Val03c,Val03e}. Recently, these scripts have been
modified to enable the discussion of conformally flat data with
non-vanishing second fundamental form. The full results of these
investigations will be presented elsewhere. Here, we shall only mention the
results which are relevant for our discussion.

Consider a conformally flat, maximal initial data set. Under the
so-called conformal Ansatz, the constraint equations reduce to:
\begin{eqnarray*}
&& \Delta\vartheta =-\frac{1}{8}\chi_{\alpha\beta}\chi^{\alpha\beta}\vartheta^{-7}, \\
&& \partial_\alpha(\vartheta^4\chi^{\alpha\beta})=0,
\end{eqnarray*}
where $\vartheta$ is the conformal factor linking with the conformal
metric,
$\delta_{\alpha\beta}=\vartheta^{-4}\widetilde{h}_{\alpha\beta}$. The
conformally rescaled second fundamental form is given by
$\chi_{\alpha\beta}=\vartheta^{-2}\widetilde{\chi}_{\alpha\beta}$. The
conformal factor $\vartheta$ can be used to compactify the initial
hypersurface $\widetilde{S}$. Let $x^\alpha$ denote normal coordinates
such that $\rho=|x|=0$ at the point at infinity of the conformally
rescaled initial hypersurface $S$.

In \cite{DaiFri01} Dain \& Friedrich have discussed a the existence of
initial data for the Einstein vacuum equations which near spatial
infinity admit expansions of the form,
\[
\widetilde{h}_{\alpha\beta}\sim \left( 1+\frac{2m}{r}\right)\delta_{\alpha\beta}+\sum_{k\geq 2}\frac{\widetilde{h}^k_{\alpha\beta}}{r^k}, \quad \widetilde{\chi}_{\alpha\beta}\sim \sum_{k\geq 2}\frac{\widetilde{\chi}^k_{\alpha\beta}}{r^k},
\]
where the coefficients involved in the expansions are smooth functions
on $S^2$. This type of data is precisely of the sort one needs in
order to use the CA scripts we have previously discussed. The
sufficient conditions they found for the existence of such data imply
that in the conformally flat case the conformally rescaled second
fundamental form has to be of the form,
\begin{equation}
\chi_{\alpha\beta}=\chi^A_{\alpha\beta}+\chi^J_{\alpha\beta}+\chi_{\alpha\beta}^Q+\O(\rho), \label{conformal_2nd}
\end{equation}
where
\begin{eqnarray*}
&& \hspace{-7mm}\chi^A_{\alpha\beta}=\frac{A}{\rho^3}\left(3n_\alpha n_\beta-\delta_{\alpha\beta}\right), \\
&& \hspace{-7mm}\chi^J_{\alpha\beta}=\frac{3}{\rho^3}\left(n_\beta\epsilon_{\gamma\alpha\rho}J^\rho n^\gamma +n_\alpha\epsilon_{\rho\beta\gamma}J^\gamma n^\rho\right), \\
&& \hspace{-7mm}\chi^Q_{\alpha\beta}=\frac{3}{2\rho^2}\left(Q_\alpha n_\beta + Q_\beta n_\alpha -(\delta_{\alpha\beta}-n_\alpha n_\beta) Q^\gamma n_\gamma\right),
\end{eqnarray*} 
and $n^\alpha=x^\alpha/\rho$ is the radial unit normal near infinity
---see also \cite{BeiOMu96}.  The constants $A$, $J^\alpha$,
$Q^\alpha$ are associated to the conformal Killing vectors on the
Euclidean 3-dimensional space generating dilatations, rotations and
``special conformal transformations'' respectively.

\medskip
For non-time symmetric data with a second fundamental form of the sort
given in equation (\ref{conformal_2nd}), the coefficients
$\phi_j^{(0)}$, $\phi_j^{(1)}$, $\phi_j^{(2)}$, and $\phi_j^{(3)}$ in
the expansion (\ref{expansion}) are smooth functions of their
arguments. However, for $\phi_j^{(4)}$ one has that,
\[
\phi_j^{(4)}=f^\infty_j + f^-_j\ln(1+\tau) +f^+_j\ln(1-\tau),
\] 
where $f^\infty_j$, $f^\pm_j$ are analytic functions of
$(\tau,\theta,\varphi)$. Furthermore, $f_j^\pm$ are polynomials in $\tau$ ---with coefficients depending on $(\theta,\varphi)$--- of the form,
\begin{subequations}
\begin{eqnarray}
&& f^+_0(\tau)=m|J|^4P_4(\tau)(1-\tau)^2, \label{f0}\\
&& f^+_1(\tau)=m|J|^3P_3(\tau)(1-\tau)^3, \label{f1}\\
&& f^+_2(\tau)=m|J|^2P_2(\tau)(1-\tau)^4, \label{f2}\\
&& f^+_3(\tau)=m|J|^2P_1(\tau)(1-\tau)^5, \label{f3}\\
&& f^+_4(\tau)=m|J|^2(1-\tau)^6,          \label{f4} 
\end{eqnarray}
\end{subequations}
and $f^-_j(\tau)=f^+_j(-\tau)$. We have also written $|J|^2=J_\alpha
J^\alpha$. The functions $P_1(\tau),\ldots,P_4(\tau)$ are polynomials
in $\tau$ such that $P_j(\pm1)\neq 0$. Thus, for conformally flat
initial data, the presence of angular momentum is incompatible with a
smooth null infinity. Note, that if the initial data is not
conformally flat ---as in the case of the standard Kerr data in
Boyer-Lindquist coordinates--- the situation is bound to be completely
different. In this case the non-conformal parts of the metric would,
in principle, leave room to cancel out the logarithmic singularities,
although this must not be necessarily the case.

\medskip
In order to compare the expansions implied by the aforementioned
results, it is convenient to change our gauge from the F-gauge to the
NP gauge. This is achieved by means of a conformal rescaling and a
rotation of the tetrad in which the conformally rescaled Weyl spinor
$\phi_{abcd}$ is expressed. A full detailed list of what are the
requirements for the NP gauge is given in \cite{FriKan00}.  After a
lengthy calculation, it can be shown that the expansions
(\ref{expansion}) together with (\ref{f0})-(\ref{f4}) imply in the NP
gauge the following expansions for the components of the Weyl spinor,
$\widetilde{\Psi}_{abcd}$, of the physical (unrescaled) spacetime,
\begin{subequations}
\begin{eqnarray}
&& \widetilde{\Psi}_0\sim k_0 m |J|^2 \Omega^3\ln \Omega + \widetilde{\Psi}_0^3\Omega^3 \cdots, \label{x0} \\
&& \widetilde{\Psi}_1\sim k_1 m |J|^2 \Omega^3\ln \Omega + \widetilde{\Psi}_1^3\Omega^3 \cdots, \label{x1} \\
&& \widetilde{\Psi}_2\sim k_2 m |J|^2 \Omega^3\ln \Omega + \widetilde{\Psi}_2^3\Omega^3\cdots,  \label{x2} \\
&& \widetilde{\Psi}_3\sim \widetilde{\Psi}^2_3\Omega^2 +\cdots, \label{x3} \\
&& \widetilde{\Psi}_4\sim \widetilde{\Psi}^1_4\Omega +\cdots.   \label{x4}
\end{eqnarray}
\end{subequations}
Here $k_0$, $k_1$ and $k_2$ denote non-zero constants, while the
coefficients $\widetilde{\Psi}_j^n$ are functions of a retarded time
$u$, which is defined up to a supertranslation, and the angular
coordinates $(\theta,\varphi)$. Thus, note that if the angular
momentum does not vanish, then null infinity not only is non-smooth,
but the \emph{Peeling behaviour} is not satisfied. It is noted that a
somewhat similar result has been obtained by Klainerman and Nicol\`{o}
---cfr. the remark of theorem 1.2 in \cite{KlaNic03}. It is also noted
that the quantity $m|J|^2$ is essentially the constants associated to
the leading logarithmic terms which have been discussed in
\cite{ChrMacSin95,Val99a}. The details of this, and a further
connection to the so-called logarithmic Newman-Penrose constants will
be also discussed elsewhere.

It is also noted that the above expansions imply that the
development of Bowen-York type data with non-vanishing ADM angular
momentum does not admit a smooth null infinity. This statement holds
for the data constructed by Brandt \& Br\"ugmann \cite{BraBru97} as
they are also conformally flat and use the same Ansatz for the second
fundamental form as the considered by Bowen \& York.

\bigskip
\emph{The stationary solutions near null infinity.-} That the
stationary solutions can be analytically extended through null
infinity is known from the appendix in \cite{DamSch90} and the results
of \cite{Dai01b}. Now, the metric given in \cite{DamSch90} is not, in
principle, in the NP gauge. Consider a null frame
$\{l^\mu,n^\mu,m^\mu,\overline{m}^\mu\}$ for that metric. A lengthy,
but straightforward argument shows that the coordinate transformations
and the Lorentz transformations required to take Damour \& Schmidt's
metric to the NP gauge are analytic functions. Indeed, one has the following,

\begin{lemma}
The stationary solutions written in the NP gauge have an analytic null
infinity. 
\end{lemma}

Moreover, it can be shown that ---see e.g. \cite{NewPen68,PenRin86}---,
\begin{subequations}
\begin{eqnarray}
&& \widetilde{\Psi}_0=\widetilde{\Psi}_0^5\Omega^5+\O(\Omega^6), \\
&& \widetilde{\Psi}_1=\widetilde{\Psi}_1^4\Omega^4+\O(\Omega^5), \\
&& \widetilde{\Psi}_2=-m \Omega^3 +\O(\Omega^4), \\
&& \widetilde{\Psi}_3=\O(\Omega^3), \\
&& \widetilde{\Psi}_4=\O(\Omega^3),
\end{eqnarray}
\end{subequations}
where the coefficients $\widetilde{\Psi}_j^n$ are $u$-independent.

\medskip
Thus, one sees that if the further requirement of stationarity is
imposed on the expansions (\ref{x0})-(\ref{x4}) then necessarily
$|J|^2=0$.

The only property of the stationary solutions
that has been used is that they admit a smooth null infinity. This may
look at first sight rather odd. However the results obtained in
\cite{Val03e,Val03c} and the ones discussed here suggest that
admitting a smooth null infinity near spatial infinity may be an
alternative characterisation of the stationary spacetimes. This,
nevertheless remains to be proved.

\acknowledgements 
I would like to thank R. Beig, M. Mars, S. Dain and
R. Lazkoz for very helpful discussions and suggestions.  This project
is funded by a Lise Meitner fellowship (M690-N09) of the Fonds zur
Forderung der Wissenschaftlichen Forschung ---FWF--- Austria. The CA
calculations described have been carried out in the computers of the
Max Planck Institut f\"ur Gravitationsphysk, Albert Einstein Institut,
Golm bei Potsdam, Germany.


\end{document}